\documentclass{article}

\usepackage{graphicx}
\usepackage{amsmath}
\usepackage{amssymb}
\usepackage{amsthm}

\newtheorem{theorem}{Theorem}
\newtheorem{lemma}{Lemma}

\theoremstyle{definition}
\newtheorem{definition}{Definition}
\newtheorem{remark}{Remark}
\newtheorem{example}{Example}
\newcommand{\Assum}{Condition}
\newtheorem{assumption}{\Assum}

\newcommand{\E}{\mathbb{E}}
\newcommand{\PP}{\mathbb{P}}

\newcommand{\A}{\mathbf{A}}
\newcommand{\B}{\mathbf{B}}
\newcommand{\D}{\mathbf{D}}
\newcommand{\I}{\mathbf{I}}
\renewcommand{\L}{\mathbf{L}}
\renewcommand{\S}{\mathbf{S}}
\newcommand{\U}{\mathbf{U}}
\newcommand{\OO}{\mathbf{O}}
\newcommand{\Q}{\mathbf{Q}}

\newcommand{\cM}{{\cal M}}
\newcommand{\cE}{{\cal E}}

\newcommand\cB{{\cal B}}
\newcommand\cV{{\cal V}}
\newcommand{\ergvar}{{\cal V}}

\newcommand{\Reals}{\mathbb{R}}

\newcommand{\spec}{\lambda}
\newcommand{\Spec}{\Lambda}

\newcommand{\efun}{u}
\newcommand{\evS}{\boldsymbol{\phi}}
\newcommand{\evU}{\boldsymbol{\psi}}

\newcommand{\obs}{\phi}
\newcommand{\Obs}{\Phi}

\newcommand{\Kdiag}{K^{\mathrm{(diag)}}(n,T)}

\newcommand{\mdef}{\stackrel{\mathrm{def}}{=}}

\newcommand{\wh}[1]{\widehat{#1}}
\newcommand{\wb}[1]{\overline{#1}}

\newcommand{\eps}{\varepsilon}

\DeclareMathOperator{\Tr}{Tr}

\newcommand{\condex}[2]{\E\left[\left. #1 \right| #2 \right]}

\begin{document}

\title{Quantum ergodicity for graphs related to interval maps}

\author{G. Berkolaiko$^1$, J.P. Keating$^2$, U. Smilansky$^3$\\
  ~\\
  $^1$ Department of Mathematics, Texas A\&M University,\\ 
  College Station, TX 77840, USA,\\
  $^2$ School of Mathematics, University of Bristol,\\ 
  Bristol BS8 1TW, UK,\\
  $^3$ Department of Physics of Complex Systems,\\ 
  Weizmann Institute of Science, Rehovot 76100, Israel}

\maketitle

\begin{abstract}
  We prove quantum ergodicity for a family of graphs that are obtained
  from ergodic one-dimensional maps of an interval using a procedure
  introduced by Pak\'onski {\em et al} ({\it J. Phys. A}, {\bf 34}, 9303-9317
  (2001)).  As observables we take the $L^2$ functions on the interval.  The
  proof is based on the periodic orbit expansion of a majorant of the quantum
  variance.  Specifically, given a one-dimensional,
  Lebesgue-measure-preserving map of an interval, we consider an increasingly
  refined sequence of partitions of the interval.  To this sequence we
  associate a sequence of graphs, whose directed edges correspond to elements
  of the partitions and on which the classical dynamics approximates the
  Perron-Frobenius operator corresponding to the map.  We show that, except
  possibly for subsequences of density 0, the eigenstates of the quantum
  graphs equidistribute in the limit of large graphs.
  
  For a smaller class of observables we also show that the Egorov property, a
  correspondence between classical and quantum evolution in the semiclassical
  limit, holds for the quantum graphs in question.
\end{abstract}

%%%%%%%%%%%%%%%%%%%%%%%%%%%%%%% SECTION %%%%%%%%%%%%%%%%%%%%%%%%%%%%%%%%%%
\section{Introduction}

The quantum ergodicity theorem is one of the central results in quantum chaos.
Essentially, it asserts that in systems in which the classical dynamics is
ergodic the probability measures associated with the squares of the moduli of
the quantum eigenfunctions converge to the classical invariant measure as one
approaches the semiclassical limit through almost all sequences of eigenstates
(any exceptional subsequences have density zero). This was originally proved
for flows \cite{Shn74,CdV85,HMR87,Zel87,GL93}, but it has since been extended
to discrete dynamical systems (chaotic maps); see, for example
\cite{DGI95,BdB96,KurRud00,KurRud01,DNW06} (for a very readable introduction
to the subject, the reader should consult \cite{DB01}).  The methods of proof
typically involve applying Egorov-type theorems, which relate the time
evolution of quantum and classical observables in the semiclassical limit.

Quantum graphs correspond to associating an operator with a graph.  For
example, this might be the discrete Laplacian acting at the vertices, or the
one-dimensional Laplacian acting on functions defined on the edges of a
(metric) graph, with matching conditions applied at the vertices.  Such
systems have recently been the subject of considerable interest
\cite{BCFK_proc}.  In particular, quantum graphs have emerged as important toy
models of quantum chaotic behaviour \cite{KS97,KS99}: if one considers
sequences of graphs with increasing numbers of edges then, under certain
conditions, the quantum eigenvalue statistics converge to those of random
matrix theory \cite{KS97,KS99,BSW02,BSW03,B04,B06,GA04,GA05}.  However,
relatively little attention has been paid to their eigenfunction statistics.
For example, quantum ergodicity has not been proved in this context.  Even
though the classical (Markovian) dynamics on a fixed graph is mixing, the
difficulty lies in dealing with sequences of graphs with increasing numbers of
bonds.  To date, the only examples that have been studied in this limit are
the star graphs (in which the bonds are connected at a single central vertex).
However, even though any given star graph is classically ergodic, the limit as
the number of bonds tends to infinity is not quantum ergodic
\cite{BKW03,BKW04,Kea06}.  This is not altogether surprising because the star
graphs do not satisfy the condition under which one expects the spectral
statistics to coincide with those of random matrix theory (instead, their
spectral statistics coincide with those of integrable systems perturbed by a
singular scatterer \cite{BK99,BBK01}).  It turns out that the star graph
eigenfunctions are strongly scarred by short periodic orbits (see also
\cite{SK03}).

The problem of finding examples of sequences of quantum graphs that are
quantum ergodic thus remains open.  It is this problem that we address here.
We start by discussing how the question of quantum ergodicity on general
graphs can be related to the ergodic properties of the eigenvectors of an {\em
  ensemble} of unitary matrices.  Each ensemble consists of matrices $\D\S_0$,
where $\S_0$ is a fixed unitary matrix, determined by the corresponding graph,
and $\D$ is a random diagonal unitary matrix.

We then identify a particular sequence of graphs (or matrices $\S_0$) for
which quantum ergodicity can be established.  These are graphs obtained from a
construction proposed by Pak\'onski {\em et al} \cite{PZK01} involving ergodic
one-dimensional maps on an interval.  We also prove the analogue of Egorov's
theorem for these graphs.

To be explicit, given a one-dimensional, Lebesgue-measure-preserving map
$S:[0,1]\to[0,1]$, we consider an increasingly refined sequence of partitions
$\cM_n$ of the interval $[0,1]$.  To this sequence we associate a sequence of
graphs $G_n$ whose directed edges (bonds) correspond to elements of the
partitions.  The quantum evolution on $G_n$ is described by a unitary
matrix $\U_n$ such that the corresponding classical (Markov) dynamics of $G_n$
approximates the Perron-Frobenius operator associated with $S$.

To a classical observable $\obs\in L^2[0,1]$ we associate a
sequence of quantum observables $\OO_n(\obs)$ which are defined as
operators corresponding to multiplication by the average value of
$\obs$ on an element of the partition. We prove that there is a sequence of
sets $J_n \subset \{1,\ldots,|\cM_n|\}$ such that
\begin{equation*}
  \lim_{n\to\infty} \frac{|J_n|}{|\cM_n|} = 1
\end{equation*}
and for all sequences $\{j_n\}_{n=1}^\infty$, $j_n\in J_n$,
\begin{equation}
  \label{eq:conv_to_mean}
  \lim_{n\to\infty}
  \left( \evU_{j_n}^{(n)}, \OO_n(\obs) \evU_{j_n}^{(n)} \right)
  = \int_0^1 \obs(x) dx,
\end{equation}
where $\evU_{j_n}^{(n)}$ is the $j_n$-th eigenvector of the graph $G_n$
(compare to the corresponding statement for cat maps, \cite{BdB96,DB01}).
This is the analogue of ``quantum ergodicity'' for the graphs in question.  It
is equivalent to the decay of the quantum variance,
\begin{equation*}
  \frac1{|\cM_n|} \sum_{j=1}^{|\cM_n|} 
  \left| \left( \evU_j^{(n)}, \OO_n(\obs) \evU_j^{(n)} \right) 
    - \int_0^1 \obs(x) dx \right|^2 \to 0
\end{equation*}
in the limit $n\to\infty$.  This equivalence follows from a Chebyshev-type
inequality and parallels the textbook proof of the statement ``for uniformly
bounded random variables, convergence in mean square is equivalent to
convergence in probability''.

If $\obs$ is Lipschitz continuous, we also prove the Egorov property,
\begin{equation*}
  \| \U_n\OO_n(\obs)\U_n^{-1} - \OO_n(\obs\circ S) \|
  = O\big(|\cM_n|^{-1}\big),
\end{equation*}
where $\U_n$ is the quantum transfer operator corresponding to the graph
$G_n$.  The existence of the Egorov property provides further justification for
the use of the term ``quantization'' when referring to the sequence $\U_n$
obtained from a map $S$.

It should be noted that we do not explicitly use the Egorov property (EP) in
the proof of quantum ergodicity (QE).  Even though the more traditional route
of deriving QE from EP is available to us, we feel that the proof in the
present form is likely to be more adaptable to other families of quantum
graphs.

This paper is organized as follows.  In Section 2 we review some of the main
issues relating to the construction of quantum graphs.  In Section 3 we
introduce the construction of Pak\'onski {\em et al} \cite{PZK01} and proceed
to discuss some of its properties.  In particular, we prove a sufficient
condition for a map to be quantizable in the fashion described by
\cite{PZK01}.  This sufficient condition, although rather restrictive,
demonstrates that the class of quantizable maps is sufficiently rich to be
interesting.

In Section 4 we introduce the observables on the quantum graphs obtained from
maps.  Their quantum variance is analyzed in Section 5.  In Sections 6 and 7 we
prove quantum ergodicity for these observables by estimating two different
contributions to the variance, and in Section 8 we prove the Egorov property
in this context.  Finally, in Section 9 we discuss some of the issues arising
in the proof of these theorems and the possibility of extending the proofs to
larger classes of graphs.

%%%%%%%%%%%%%%%%%%%%%%%%%%%%%%% SECTION %%%%%%%%%%%%%%%%%%%%%%%%%%%%%%%%%%
\section{Quantum graphs}
\label{sec:q_graphs}

A {\it quantum graph} can be defined in two different, but
related, ways.  In both constructions we start with a graph $G =
(\cV,\cB)$ where $\cV$ is a finite set of vertices (or nodes), and
$\cB$ is the set of bonds (or edges).  Each bond $b$ has a
non-zero length, denoted $L_b$.  The lengths $L_b$ are assumed to
be rationally independent.

The first way to define a quantum graph \cite{KS99} is to identify each bond
$b$ with the interval $[0,L_b]$ of the real line and thus define the
$L^2$-space of functions on the graph.  Then one can consider the eigenproblem
\begin{equation}
  \label{eq:shrod_eq}
  -\frac{d^2}{d x^2} \efun_b(x) = \spec^2\efun_b(x).
\end{equation}
This setup has been studied by mathematicians since the 1980s
\cite{Lum80,Roth83b,vB85,Nic87,PP88} and was used in physical models prior to
that \cite{Pau36,Gri53,RS53}.

To make the operator in (\ref{eq:shrod_eq}) self-adjoint we need to impose
matching conditions on the behavior of $\efun$ at the vertices of the graph.
One possibility is to impose Kirchhoff conditions:\footnote{sometimes called
  ``Neumann'' conditions} we require that $\efun$ is continuous on the
vertices, and that the probability current is conserved, i.e.
\begin{equation}
  \label{eq:Neu2}
  \sum_{v \in b} \frac{d}{d x} \efun_b(v) = 0 \qquad
  \mbox{ for all } v\in\cV,
\end{equation}
where the sum is over all bonds that originate from the vertex $v$ (the bonds
are now taken to be undirected) and the derivatives are taken at the vertex $v$
in the outer direction.  The admissible boundary conditions were classified
in, among other sources, \cite{KosSch99, Har00}.

The second construction considers wave propagation on the graph where each
vertex is treated as a scatterer and the propagation along the bonds is free.
This construction was first considered in \cite{KS99} and generalized in
\cite{Tan00} to directed graphs.

In both constructions one ends up with a unitary matrix $\S(\spec) =
e^{i\spec\L}\S(0)$, where $\L$ is the diagonal matrix of the bond
lengths.  This matrix gives the eigenvalues $\{\spec_n\}$ of
(\ref{eq:shrod_eq}) via the equation
\begin{equation}
  \label{eq:sec_eq}
  \det (\I - \S(\spec_n) ) = 0.
\end{equation}
The dimension of the above matrices is equal to the number $B$ of {\em
  directed} bonds of the graph $G$.  If the bonds were initially undirected,
each bond is split into two directed bonds of the same length.

In various sources the notion of the ``spectrum $\sigma(G)$ of the graph $G$''
can refer either to the eigenproblem (\ref{eq:shrod_eq}) (and thus solutions
of (\ref{eq:sec_eq})) or to the eigenphases of the matrix $\S(\spec)$ for an
arbitrary $\spec$.  This is not as confusing as it might seem, since the
statistical properties of both versions of the spectrum are conjectured to
coincide when averaged over a large interval of $\spec$.

Similarly, the ``eigenvector'' of $G$ can refer to one of three
objects:
\begin{enumerate}
\item the function $\efun(x)$ that solves (\ref{eq:shrod_eq}), subject to
  boundary conditions, for some $\spec\in\sigma(G)$,
\item the eigenvector of $\S(\spec_n)$ corresponding to the eigenvalue 1,
  denoted by $\evS_n$,
\item any eigenvector of $\S(\spec)$ for arbitrary $\spec$, denoted by
  $\evU(\spec)$.
\end{enumerate}
There is a simple correspondence between the first two notions of the
eigenvector: the solution $\efun(x)$ is a superposition of plane waves with
coefficients given by the elements of $\evS_n$.  Below we discuss a heuristic
formula which connects the ergodic properties of the second and the third
types of eigenvectors.  This formula provides an additional motivation for the
results in the main body of our paper, where we study the eigenvectors
$\evU(\spec)$.  It should be mentioned that these results are fully rigorous
and do not rely on the heuristic connection.

To proceed, we need to introduce more notation.  By $\evU_k(\spec)$ we will
denote the $k$-th eigenvector of $\S(\spec)$.  Our observables are diagonal
matrices $\OO$ acting in the space of directed bonds.  The matrix $\L$, as
before, is the diagonal matrix of the bond lengths.  The average bond length,
$B^{-1}\Tr\L$, is denoted by $\bar{L}$.  Quantum ergodicity is the property of
almost all eigenvectors to equidistribute.  This is equivalent to the
vanishing of the variance in some limit.  For example, we would like to prove
that the variance of $\langle \evS_n | \OO | \evS_n \rangle - B^{-1}\Tr\OO$
(and, correspondingly, $\langle \evU_k(\spec) | \OO | \evU_k(\spec) \rangle -
B^{-1}\Tr\OO$) vanishes.  At this point two obvious questions arise: (a) with
respect to which ensemble is the variance taken, and (b) in which limit is it
expected to vanish?

Taking, without loss of generality, $\Tr\OO$ to be zero, we define two
variances
\begin{eqnarray*}
  V^S(\Spec,B) &=& \frac1{\overline{N}(\Spec)} \sum_{\spec_n \leq \Spec}
  \langle \evS_n|\OO|\evS_n\rangle^2,\\
  V^U(\S(\spec),B) &=& 
  \frac1{B} \sum_{k=1}^B \langle \evU_k(\spec)|\OO|\evU_k(\spec)\rangle^2,
\end{eqnarray*}
where $\overline{N}(\Spec) = \Spec \Tr\L / 2\pi$ is the mean number of the
eigenvalues in the interval $[0,\Spec]$.

A heuristic calculation presented in Appendix~\ref{sec:var_rel} suggests that,
if the bond lengths are rationally independent,
\begin{equation}
  \label{eq:var_rel}
  \lim_{\Spec\to\infty} \frac1{\overline{N}(\Spec)} \sum_{\spec_n\leq\Spec}
  \frac{\langle \evS_n|\OO|\evS_n\rangle^2} 
  {\langle \evS_n | \L | \evS_n \rangle / \bar{L}}
  = \lim_{\Spec\to\infty} \frac1\Spec \int_0^\Spec V^U(\S(\spec),B) d\spec.
\end{equation}
Thus, if the lengths of the bonds are taken from a narrow distribution, the
two variances are intimately connected.  Moreover, following \cite{BaGa00} one
can show that the limit on the right-hand side coincides with the average of
$V^U(\D\S(0),B)$, where $\D$ are uniformly distributed random unitary diagonal
matrices.  Thus equation~(\ref{eq:var_rel}) relates the quantum ergodic
properties of a graph to the like properties of an {\em ensemble\/} of random
matrices.

Equation~(\ref{eq:var_rel}) suggests that one cannot in general expect the
variance to vanish in the limit $\spec\to\infty$.  It is natural, however, to
expect ergodicity in the limit $B\to\infty$ (cf.\ \cite{BdB96}).  A serious
associated problem here is the choice of an appropriate sequence of graphs and
observables.  One sequence of graphs, the quantum star graphs, has been
investigated in \cite{BKW03,BKW04}, and it was found, in particular, that the
variance $V^S(\Spec,B)$ does not vanish even when $B\to\infty$.  This is not
altogether surprising because the star graphs are known to exhibit
non-standard spectral statistics \cite{KS99,BK99}, corresponding to integrable
systems perturbed by a point-scatterer, rather than to chaotic systems
\cite{BBK01} (for a review of the quantum fluctuation statistics of star
graphs see \cite{Kea06}).  This is due to the fact that the spectral gap in
their Markov transition matrix closes more quickly as $B\to\infty$ (like
$1/B$) than is the case for graphs exhibiting truly quantum chaotic behaviour.
The lack of quantum ergodicity for the star graphs is related to the existence
of strong scarring of the eigenfunctions by short periodic orbits.  In the
present article we study sequences of graphs generated from 1-dimensional maps
of an interval in a fashion suggested in \cite{PZK01}.  We prove that for a
suitable choice of observables, the variance $V^U(\S(\spec),B)$ converges to 0
for any $\spec$, given that the original 1-dimensional map was ergodic.  This
is a stronger statement than the convergence when averaged with respect to
$\spec$, as suggested by relation~(\ref{eq:var_rel}).

%%%%%%%%%%%%%%%%%%%%%%%%%%%% New Section %%%%%%%%%%%%%%%%%%%%%%%%%%%%%%
\section{Quantum graphs obtained from 1d maps}

Pak\'onski {\em et al} \cite{PZK01} proposed a procedure to associate a
sequence of quantum graphs to a one-dimensional map of an interval.  In this
section we review their construction and proceed to investigate some of its
properties.

We consider maps of an interval, which we take to be $[0,1]$.  A {\em
  partition} $\cM$ of the interval $[0,1]$ will be taken to mean a finite
collection of open disjoint intervals $E_j$ (henceforth called {\em atoms})
such that 
\begin{equation*}
  [0,1] = \overline{\bigcup_{j=1}^M E_j},
\end{equation*}
where $M = |\cM|$ denotes the number of intervals in the partition.  We will
denote by $\cE(\cM)$ the set of endpoints of the partition $\cM$.  In a slight
abuse of the notation we will also denote by $\cM$ the $\sigma$-algebra
generated by the atoms of the partition $\cM$.  When considering sequences
$\{\cM_n\}$ of partitions, each partition will be a refinement of the previous
one, $\cE(\cM_n)\subset\cE(\cM_{n+1})$.  We will write $\cM_n \subseteq
\cM_{n+1}$ to describe this statement.

\begin{assumption}
  \label{assum:map}
  \renewcommand{\theenumi}{(\alph{enumi})}
  \renewcommand{\labelenumi}{\theenumi} 
  We consider maps $S:[0,1]\to[0,1]$ that satisfy the following conditions:
  \begin{enumerate}
  \item \label{item:a_map_meas_pres} the Lebesgue (uniform) measure $\mu$ is
    preserved by the map $S$: $\mu(A) = \mu\left(S^{-1}(A)\right)$ for any
    measurable set $A$;
  \item \label{item:a_map_linear} there exists a partition $\cM_0$ of the
    interval $[0,1]$ into $M_0$ {\em equal\/} atoms, with $S$ linear on each
    atom;
  \item \label{item:a_map_endpoints} the set of endpoints $\cE(\cM)$ is
    forward-invariant under the action of $S$: $S(\cE)\subset \cE$.
  \end{enumerate}
\end{assumption}

\begin{figure}[t]
  \centering
  \includegraphics[scale=0.8]{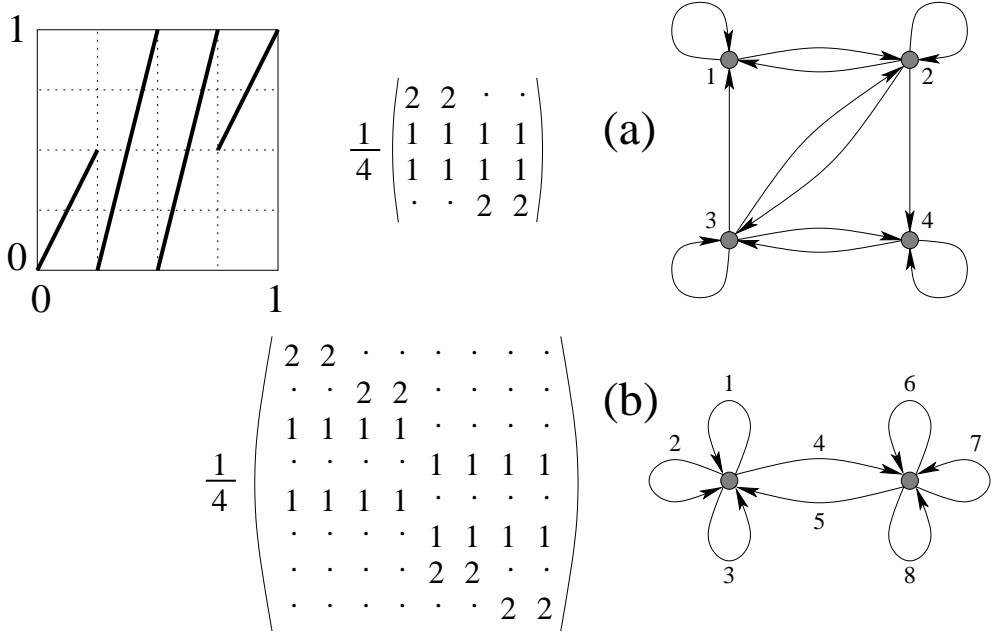}
  \caption{An example of a quantizable map and the corresponding matrices
    $\B$: (a) $|\cM|=4$, the atoms of the partition are represented by the
    {\em vertices\/} of the graph; (b) $|\cM|=8$, the atoms of the partition
    are represented by the {\em edges\/} of the graph.}
  \label{fig:map1}
\end{figure}

An example of a map $S$ satisfying \Assum~\ref{assum:map} is shown on
Figure~\ref{fig:map1}.  The tent map with slope 2 is another such example.

\begin{remark}
  \label{rem:atom_onto}
  The conditions on the map $S$ imply, in particular, that for any two atoms
  $E$ and $E'$, either $S(E)$ is disjoint with $E'$ or $S(E) \supset E'$.
\end{remark}

\begin{remark}
  It is possible to generalize the construction to maps that preserve a
  measure different from Lebesgue, but such a map would have to be
  topologically conjugate to a piecewise linear map satisfying the above
  properties.  To maintain a degree of generality we strive to make explicit
  the conditions that are imposed on the map $S$ and the measure $\mu$.
\end{remark}

The Frobenius-Perron operator, reduced to measures constant on
each atom of a partition $\cM$, can be described by a matrix $\B$
of size $M=|\cM|$.  The entries of the matrix are given by
\begin{equation} \label{eq:markov_def}
  B_{jk} = \frac{\mu\left(E_j\cap S^{-1}(E_k)\right)}{\mu(E_j)},
\end{equation}
and can be described as the answer to the question ``what proportion of the
set $E_j$ gets mapped into $E_k$''.  An example of a map $S$ and the
corresponding matrices $\B$ for two different partitions are shown on
Figure~\ref{fig:map1}.

If we view the interval $[0,1]$ with the uniform measure as a probability
space, we can write $B_{jk} = \PP\left(S(x)\in E_k | x\in E_j\right)$.

\begin{lemma}
  \label{lem:B_properties}
  Let the set of endpoints of the partition $\cM$ be invariant under $S$.
  Then the matrix $\B$ defined by (\ref{eq:markov_def}) satisfies the
  following properties:
  \begin{enumerate}
  \item \label{item:B_stoch} $\B$ is stochastic,
    \begin{equation*}
      \sum_{k=1}^M B_{jk} = 1.
    \end{equation*}
  \item \label{item:B_doub_stoch} If the atoms $E_j$ of the partition $\cM$
    have equal measure and the map $S$ preserves this measure then $\B$ is
    doubly stochastic
    \begin{equation*}
      \sum_{j=1}^M B_{jk} = 1.
    \end{equation*}
  \item \label{item:prod_B_measure} If the atoms $E_j$ of the partition $\cM$
    have equal measure and if the map $S$ is linear with respect to $\mu$ on
    each atom $E_j$ (i.e. $\mu(S(A))=C\mu(A)$ for some $C$ and all $A\in E_j$)
    then
    \begin{equation}
      \label{eq:prod_B_measure}
      B_{j_0j_1}B_{j_1j_2}\cdots B_{j_{k-1}j_k}
      = \frac{\mu\left( \bigcap_{r=0}^k S^{-r}\left(E_{j_r}\right)\right)}
      {\mu\left(E_{j_0}\right)}.
    \end{equation}
  \end{enumerate}
\end{lemma}

\begin{proof}
  Part \ref{item:B_stoch} follows directly from (\ref{eq:markov_def}),
  \begin{equation*}
    \sum_{k=1}^M B_{jk} = \frac{1}{\mu(E_j)} 
    \,\,\mu\!\left(E_j \cap \left(\bigcup_{k=1}^M S^{-1}(E_k)\right)\right)
    = \frac{\mu\left(E_j\cap [0,1]\right)}{\mu(E_j)} = 1.
  \end{equation*}

  Part \ref{item:B_doub_stoch} is similar: if $\mu(E_j)=m$ for all $j$ then
  \begin{equation*}
    \sum_{j=1}^M B_{jk} = \frac1m 
    \,\,\mu\!\left( \left(\bigcup_{j=1}^M E_j \right) \cap S^{-1}(E_k)\right)
    = \frac{\mu\left( S^{-1}(E_k) \right)}{m} = \frac{\mu(E_k)}{m} = 1.
  \end{equation*}
  
  Part \ref{item:prod_B_measure} is a consequence of the fact that, if
  $\mu(S(A))=C_j\mu(A)$ for all $A\subset E_j$, then $B_{jk}$ is either $0$ or
  $1/C_j$.  Consider first the case $B_{j_{r}j_{r+1}} = 0$ for some $r$.  By
  definition of $\B$, this means that $\mu\left(E_{j_{r}} \cap
    S^{-1}(E_{j_{r+1}})\right) = 0$.  Therefore,
  \begin{equation*}
    \mu\left(S^{-r}(E_{j_{r}}) \cap S^{-r-1}(E_{j_{r+1}})\right) = 0,
  \end{equation*}
  and the
  expression on the right hand side of (\ref{eq:prod_B_measure}) evaluates to
  zero. 
  
  Now consider the case $n=2$ (the case of general $n$ being analogous) with
  both $B_{j_0 j_1}$ and $B_{j_1j_2}$ being different from zero.  Let, to
  simplify the notation, $j_r=r$.  Then
  \begin{eqnarray*}
    B_{0,1} = \frac{\mu\left(E_0\cap S^{-1}(E_1)\right)}{\mu(E_0)}
    &=& \frac{\mu\left(E_0\cap S^{-1}(E_1)\right)}{\mu(E_1)} \\
    &=& \frac{\mu\left(E_0\cap S^{-1}\left(E_1\cap S^{-1}(E_2)\right)\right)}
    {\mu\left(E_1\cap S^{-1}(E_2)\right)},
  \end{eqnarray*}
  where the last equality is true by virtue of linearity of $S$.  Using the
  definition of $B_{1,2}$ and the identity $S^{-1}\left(A \cap B\right) =
  S^{-1}\left(A\right) \cap S^{-1}\left(B\right)$, we arrive to
  \begin{equation*}
    B_{0,1}B_{1,2} 
    = \frac{\mu\left(E_0\cap S^{-1}(E_1) \cap S^{-2}(E_2)\right)}{\mu(E_1)},
  \end{equation*}
  which is the sought result, given that $\mu(E_1) = \mu(E_0)$.
\end{proof}

As mentioned earlier, we are interested in sequences of partitions.

\begin{assumption}
  \label{assum:partition}
  \renewcommand{\theenumi}{(\alph{enumi})}
  \renewcommand{\labelenumi}{\theenumi}
  We consider sequences of partitions $\cM_n$ that satisfy
  \begin{enumerate}
  \item the atoms within each of the partition have equal measure;
  \item the sets of endpoints $\cE(\cM_n)$ are forward-invariant under the
    action of $S$;
  \item \label{item:a_part_preimage} the set of the endpoints of $\cM_n$
    contains the $j$-th pre-image of the endpoints of $\cM_0$ for all
    $j=1,\ldots n$.
  \end{enumerate}
\end{assumption}

\begin{remark}
  Given a map $S$ satisfying \Assum~\ref{assum:map} one can always
  construct a sequence of partitions satisfying
  \Assum~\ref{assum:partition}.
\end{remark}

\begin{remark}
  \Assum{s}~\ref{assum:map}\ref{item:a_map_linear} and
  \ref{assum:map}\ref{item:a_map_endpoints} imply that the map $S$ is
  non-contracting, $\mu(S(A)) \geq \mu(A)$.  If the map is ergodic (see
  definition~\ref{def:ergodicity} in section~\ref{sec:quant_observable}),
  \Assum~\ref{assum:map}\ref{item:a_map_meas_pres} implies that the preimages
  of $\cE(\cM_0)$ with respect to $S$ are dense in $[0,1]$.  This, in turn,
  implies that the size of the atoms of the partitions $\cM_n$ tends to zero
  (or, equivalently, $M_n\to\infty$).
\end{remark}

The following lemma explains the way in which such sequences of partitions
`resolve' the dynamics.

\begin{lemma}
  \label{lem:traj_spec}
  Given a partition $\cM_n$ satisfying \Assum~\ref{assum:partition}, let $k_0$
  and $k_n$ be such that $S^n(E_{k_0}) \supset E_{k_n}$ (cf.\ 
  Remark~\ref{rem:atom_onto}).  Then there exists a unique sequence $k_1,
  \ldots, k_{n-1}$ such that
  \begin{equation*}
    x\in E_{k_0} \mbox{ and } S^n(x)\in E_{k_n} 
    \qquad \Rightarrow \qquad
    S^j(x) \in E_{k_j} \quad \forall 0\leq j\leq n.
  \end{equation*}
\end{lemma}

\begin{proof}
  Consider an atom $E$ of the partition $\cM_n$.
  \Assum~\ref{assum:partition}\ref{item:a_part_preimage} means that
  for every $j=0,\ldots k$, the image $S^j(E)$ lies in a single atom of the
  ``primary'' partition $\cM_0$.  Since the map $S$ is one-to-one on each atom
  of $\cM_0$, we conclude, by induction, that $S^j$ is one-to-one on $E$ for
  every $j=1,\ldots n+1$.
  
  Assume that the statement of the lemma is incorrect: there are two points,
  $x$ and $y$, that satisfy, without loss of generality, $x,y\in E_1$,
  $S^r(x)\in E_2$, $S^r(y)\in E_3$ and $S^n(x), S^n(y)\in E_4$.
  Remark~\ref{rem:atom_onto} implies the following inclusions:
  \begin{equation*}
    S^r(E_1) \supset E_2, \qquad S^2(E_1)\supset E_3, \qquad
    S^{n-r}(E_2) \supset E_4, \qquad S^{n-r}(E_3) \supset E_4.
  \end{equation*}
  Thus each $z\in E_4$ has $n-r$-preimages in both sets $E_2$ and $E_3$ and,
  therefore, two distinct $n$-preimages in $E_1$.  This contradicts the
  earlier conclusion that $S^n$ is one-to-one on $E_1$.
\end{proof}

\begin{remark}
  Obviously, Lemma~\ref{lem:traj_spec} is valid if, instead of the
  ``position'' of $S^n(x)$ (i.e. the atom $E_j$ such that $S^n(x)\in E_j$), we
  know the position of $S^m(x)$ for some $m<n$: we can still recover positions
  of all iterates $S^j(x)$ for $0<j<m$.  In fact, a careful inspection of the
  proof reveals that the Lemma would still be true for $m=n+1$.  However, if
  we know only that $x\in E_1$ and $S^{n+2}(x)\in E_2$, we would not be able
  to pinpoint $S^j(x)$, $0<j<n+2$, to any particular atom of the partition
  $\cM_n$.
\end{remark}

The next lemma exhibits the block structure of the matrix $\B$.

\begin{lemma}
  \label{lem:intersect_classes}
  For a partition $\cM_n$, $n>0$, define an equivalence relation between atoms
  by setting $E_j \sim E_k$ if $S(E_j)$ intersects $S(E_k)$ and then
  completing by transitivity.  Then the maximum number of elements in an
  equivalence class is uniformly bounded with respect to $n$.
\end{lemma}

For example, in the partition of Figure~\ref{fig:map1}, part (b), the atoms
$E_1$, $E_2$, $E_3$ and $E_5$ form one equivalence class and the other four
atoms form another equivalence class.  Note that, if the atoms of a partition
are represented by edges of the graph, the equivalence classes correspond to
the groups of edges ending in the same vertex.  For the map in
Figure~\ref{fig:map1} the uniform bound on the size of an equivalence class is
4, as will be evident from the proof.

\begin{proof}
  Take an atom $A$ of the primary partition $\cM_0$ and let $(x_1,y_1),\ldots,
  (x_k,y_k)$ be the disjoint intervals forming the pre-image of $A$ with
  respect to $S$.  By \Assum~\ref{assum:map}\ref{item:a_map_endpoints} these
  intervals contain no endpoints of $\cM_0$, therefore the map $S$ is linear
  on each interval.  \Assum~\ref{assum:map} also implies that all slopes
  of the map $S$ are integer.  Denote the slope of $S$ on the interval
  $(x_j,y_j)$ by $s_j$.  To simplify the notation we assume that all $s_j$ are
  positive.  Let $p$ be the least common multiple of $s_j$.
  
  \Assum~\ref{assum:partition}\ref{item:a_part_preimage} implies that $x_j$
  and $y_j$ are endpoints of the partition $\cM_n$.  Choose
  $x_j'\in\cE(\cM_n)$ such that the interval $(x_j,x_j')$ contains exactly
  $p/s_j$ atoms of the partition $\cM_n$.  Since the atoms of $\cM_n$ have
  equal length (which we denote by $\mu_n$),
  \begin{equation}
    \label{eq:blown_lengths_equal}
    s_{j_1}(x_{j_1}'-x_{j_1}) = s_{j_2}(x_{j_2}'-x_{j_2}) = p\mu_n,\qquad
    \mbox{for any }j_1,j_2.
  \end{equation}
  Moreover, the selected points $x_j'$ are the {\em closest\/} to the
  respective $x_j$ to satisfy both condition (\ref{eq:blown_lengths_equal})
  and $x_j'\in\cE(\cM_n)$.  In particular, this implies that
  $x_j'\leq y_j$, since setting $x_j'=y_j$ would also satisfy
  condition~(\ref{eq:blown_lengths_equal}).

  From the above we can conclude that $S$ maps all intervals $(x_j,x_j')$ to
  the same subinterval of $A$.  The atoms of $\cM_n$ making up the intervals
  $(x_j,x_j')$ thus form an equivalence class of size $p/s_1+\cdots+p/s_k$,
  which is independent of $n$.
  
  We can now repeat this procedure with intervals $(x_1',y_1),\ldots,
  (x_k',y_k)$ and, thereafter, with all atoms $A$ of the partition $\cM_0$.
  If some of the slopes $s_j$ are negative, the procedure would still go
  through with minor variations.
\end{proof}

Having obtained a sequence of doubly stochastic matrices $\B_n$ we define
their ``quantizations'' as unitary matrices $\U_n$ such that
\begin{equation}
  \label{eq:U_def}
  (\B_n)_{jk} = \left| (\U_n)_{jk} \right|^2.
\end{equation}
The doubly stochastic matrices $\B$ for which finding a corresponding $\U$ is
possible are called {\em unistochastic}.

\begin{assumption}
  \label{assum:unistoch}
  We assume that the map $S$ is such that all of the corresponding matrices
  $\B_n$, bar finitely many, are unistochastic.
\end{assumption}

Not all bistochastic matrices are unistochastic.  However, formulating a
general sufficient conditions that ensure unistochasticity is a question of
considerable difficulty.  The interested reader is referred to
\cite{PZK01,ZKSS03} and the references therein where some necessary conditions
are discussed and where examples of maps satisfying and failing
\Assum~\ref{assum:unistoch} are given.  To convince the reader that the class
of maps $S$ satisfying \Assum~\ref{assum:unistoch} is far from empty we state
the following {\em sufficient\/} condition.

\begin{lemma}
  \label{lem:equal_slopes}
  If the slopes of the map $S$ satisfying \Assum~\ref{assum:map} are all equal
  (modulo sign), \Assum~\ref{assum:unistoch} is also satisfied.
\end{lemma}

\begin{proof}
  This Lemma follows simply from the proof of
  Lemma~\ref{lem:intersect_classes}.  Indeed, let $s$ be the absolute value of
  the slope of $S$.  Then all matrices $\B_n$, $n>0$, have a block structure
  with blocks of the size $s\times s$ and elements $1/s$.  Thus the question
  is really about finding an $s\times s$ unitary matrix with all elements
  satisfying $|U_{jk}|^2=1/s$.  One example of such matrix is the Fourier
  matrix with elements $U_{jk} = \exp\{2\pi i jk/s\}/\sqrt{s}$.
\end{proof}

\begin{example}
  An example of a map $S$ which has unequal slopes but is still unistochastic
  is provided by the map of Figure~\ref{fig:map1}.
\end{example}

\begin{remark}
  An observant reader would notice that, given one unitary $\U$ satisfying
  (\ref{eq:U_def}), one can produce infinitely many such matrices.  For
  example, one can multiply a given $\U$ by an arbitrary diagonal unitary
  matrix.  However, the results of our paper do not depend on the precise
  choice of matrices $\U_n$, provided that condition~(\ref{eq:U_def}) is
  satisfied.
\end{remark}

One can associate a graph to the matrices $\B$ and $\U$ in the following way:
the indices of the matrices enumerate the {\em directed edges\/} of the graph;
the end of an edge $j$ coincides with the start of the edge $k$ if the matrix
element $B_{jk}$ is non-zero.  The number of distinct vertices in such a
construction should be maximized, then the vertices will correspond to the
equivalence classes of Lemma~\ref{lem:intersect_classes}.

The matrix $\B$ defines a Markov chain on the edges of the graph with $B_{jk}$
representing the transition probability from $j$ to $k$.  The matrix $\U$ can
be viewed as a quantum propagator on the graph.  This geometrical
interpretation of the two matrices as a graph will be helpful in the later
sections when we use trajectories on the graph to describe properties of the
eigenvectors of $\U$.

It is also possible to associate {\em vertices\/} of a graph to the indices of
$\B$, see Figure~\ref{fig:map1}, part (a).  We use directed edges for
reasons of tradition, rather than convenience.

%%%%%%%%%%%%%%%%%%%%%%%%%%%% New Section %%%%%%%%%%%%%%%%%%%%%%%%%%%%%%
\section{Quantization of the observables}
\label{sec:quant_observable}

Having defined the sequences of unitary matrices $\U_n$, ergodic properties of
whose eigenvectors we are going to study, we need a final ingredient, the
observables $\OO_n$.  For a general sequence of graphs, it is not obvious how
to define a consistent sequence of observables.  In our case, however, there
is a natural answer.

We use the discretizations of functions $\obs\in L^2[0,1]$ as our observables.
Fix a partition $\cM$ (the semi-classical limit corresponds to
$|\cM|\to\infty$).  If the function $\obs$ is constant on each atom of the
partition $\cM$, its quantization $\OO = \OO(\obs)$ is a diagonal matrix with
entries $O_{jj} = \obs(x)$ where $x\in E_j$.  If $\obs$ is not constant on the
atoms of $\cM$, we replace $\obs$ by its local average.  More precisely, we
introduce the piecewise constant function $\wh{\obs}$ defined by
\begin{equation*}
  \wh{\obs}(x) = \frac1{\mu(E_j)} \int_{E_j} \obs(y) d\mu(y), \qquad
  \mbox{where }E_j \ni x.
\end{equation*}
Then we define $\OO = \OO(\phi)$ as before, by
\begin{equation}
  \label{eq:O_def}
  O_{jj} = O_{jj}(\obs) = \wh{\obs}(E_j)
  = \frac1{\mu(E_j)} \int_{E_j} \obs(y) d\mu(y).
\end{equation}

It is convenient to describe $\wh{\obs}$ using the notions of
probability theory.  In probabilistic language, $\obs$ is a random
variable defined on the probability space $[0,1]$ and $\wh{\obs}$
is its conditional expectation, $\wh{\obs} = \condex{\obs}{\cM}$.
We will also use the notation of expectation to denote the integral over
$[0,1]$:
\begin{equation*}
  \E\obs = \int_0^1 \obs(x) d\mu(x).
\end{equation*}
In particular, $\|\obs\|_2 = \left(\E\obs^2\right)^{1/2}$.

To prove quantum ergodicity, we will rely on the ergodicity of the classical
map $S$.  Since our observables are in $L^2$, the relevant version of the
ergodic theorem is the $L^2$ ergodic theorem (see, e.g., \cite{billingsley}).
\begin{definition}
  \label{def:ergodicity}
  A map $S:[0,1]\to[0,1]$ is {\em ergodic\/} if any set $A \subset [0,1]$
  satisfying
  \begin{equation*}
    S^{-1}(A) = A
  \end{equation*}
  has either full or zero measure.
\end{definition}
\begin{theorem}{\bf ($L^2$ Ergodic Theorem)}
  \label{thm:L2ergodic}
  If $\obs\in L^2[0,1]$ and $S$ is ergodic, then
  \begin{equation} \label{eq:l2-erg}
    \ergvar_T(\obs) 
    \mdef \E\left(\frac1T\sum_{t=1}^T\obs\circ S^t - \E\obs\right)^2 \to 0.
  \end{equation}
\end{theorem}

Since the ergodic theorem applies to any function from $\obs\in L^2$, it also
applies to $\wh{\obs}$, whatever partition $\cM$ was used to produce it.
Unfortunately, a uniform estimate for the rate of convergence in
(\ref{eq:l2-erg}) for different hat-versions of the same $\obs$ is not known
\cite{Kac96}.  However, it is easy to see that, for fixed $T$,
\begin{equation}
  \label{eq:local_av_conv}
  \ergvar_T(\wh{\obs}) \to \ergvar_T(\obs)
\end{equation}
as the partition in the definition of $\wh{\obs}$ gets finer.

%%%%%%%%%%%%%%%%%%%%%%%%%%%%%%% New Section %%%%%%%%%%%%%%%%%%%%%%%%%%%%%%%
\section{Quantum variance of an observable}

Given a map $S$ and a sequence of partitions $\cM_n$ we have constructed a
sequence of Markov matrices $\B_n$, which, in turn, give rise to unitary
matrices $\U_n$.  On the other hand we are given an observable $\obs$ and we
have constructed a corresponding sequence of diagonal matrices $\OO_n$, which
``quantize'' $\obs$.  We denote by $M_n$ the number of atoms in the partition
$\cM_n$.  This is also the size of the matrices $\B_n$, $\U_n$ and $\OO_n$.
The semiclassical limit corresponds to $M_n\to\infty$.

Let $\evU^{(n)}_j$, where $j=1,\ldots,M_n$ denote the orthonormal eigenvectors
of $\U_n$.  If an eigenvalue is degenerate, the particular choice of the basis
for its eigenspace is unimportant.  As discussed earlier, to show quantum
ergodicity it is sufficient to prove that
\begin{equation}
  \label{eq:quantum_mean}
  \frac1{M_n} \sum_{j=1}^{M_n} \left(\evU^{(n)}_j, \OO_n \evU^{(n)}_j\right)
  \to \E(\obs) = \int_0^1 \obs(x) dx
\end{equation}
and
\begin{equation}
  \label{eq:quantum_variance}
  V_n = \frac1{M_n} \sum_{j=1}^{M_n}
  \left| \left(\evU^{(n)}_j, \OO_n \evU^{(n)}_j\right) - \E(\obs) \right|^2
  \to 0,
\end{equation}
as $n\to\infty$.  It is straightforward to verify (\ref{eq:quantum_mean}).
Indeed, from the unitarity of $\U_n$ and the definition of $\OO_n$,
\begin{equation*}
  \frac1{M_n} \sum_{j=1}^{M_n} \left(\evU^{(n)}_j, \OO_n \evU^{(n)}_j\right)
  = \frac1{M_n} \Tr\OO_n = \E\left(\wh{\obs}\right)
  = \E(\obs).
\end{equation*}
Thus the main task is to show (\ref{eq:quantum_variance}).

Without loss of generality we can assume that $\E(\obs)=0$.  In
what follows we will omit the sub- and super-scripts $n$ unless we
want to underline the dependence of a quantity on $n$ and on the
partition $\cM_n$.

To obtain an estimate of $V_n$ we employ some standard
manipulations. If $\evU$ is an eigenvector of a unitary matrix
$\U$, we have, for any matrix $\OO$ (not necessarily diagonal) and
all $t\in\mathbb{N}$
\begin{equation*}
  \left(\evU, \OO \evU\right) = \left(\U^t\evU, \OO \U^t\evU\right)
  = \left(\evU, (\U^*)^t\OO \U^t\evU\right).
\end{equation*}
Summing this equality over $t=0,\ldots T-1$ we obtain
\begin{equation*}
  \left(\evU, \OO \evU\right)
  = \left(\evU, \frac1T \sum_{t=0}^{T-1}(\U^*)^t\OO \U^t\evU\right).
\end{equation*}
We introduce the shorthand $\OO_{n,T}$ for the time average of $\OO_n$,
\begin{equation*}
  \OO_{n,T} = \frac1T \sum_{t=0}^{T-1}(\U_n^*)^t\OO_n \U_n^t.
\end{equation*}
Using Cauchy-Schwarz inequality and orthonormality of $\{\evU_j\}$
we estimate
\begin{equation*}
  \left|\left(\evU, \OO \evU\right)\right|^2
  = \left|\left(\evU, \OO_T \evU\right)\right|^2
  \leq \left(\OO_T \evU, \OO_T \evU\right) 
  = \left(\evU, \OO_T^*\OO_T \evU\right),
\end{equation*}
and obtain
\begin{eqnarray}
  \nonumber
  V_n &=& \frac1{M_n} \sum_{j=1}^{M_n}
  \left| \left(\evU_j, \OO \evU_j\right) \right|^2 \\
  \label{eq:K_def}
  &\leq& \frac1{M_n} \sum_{j=1}^{M_n} \left(\evU_j, \OO_T^*\OO_T \evU_j\right)
  = \frac1{M_n} \Tr \left(\OO_T^*\OO_T\right) \mdef K(n,T)
\end{eqnarray}

It is important to note that the above inequality is valid {\em for all\/}
values of $T$.  Thus, to show that $V_n\to0$, we are free to
choose an appropriate $T=T(n)$ for each $n$ as long as we can demonstrate that
\begin{equation*}
  K(n,T(n)) = \frac1{M_n}\Tr\left(\OO_{n,T(n)}^*\OO_{n,T(n)}\right) \to 0,
\end{equation*}
as $n\to\infty$.  In the following sections we prove that $T(n)=n$ is a
suitable choice for this task.

For our purposes, it is more convenient to work with the matrices $\S_{n,T}$
defined by
\begin{equation*}
  \S_{n,T} = \U_n^T\OO_{n,T} = \frac1T \sum_{t=0}^{T-1}\U_n^{T-t}\OO_n \U_n^t,
\end{equation*}
which is equivalent to working with $\OO_T$ since
$\S^*_T\S_T = (\U^T\OO_T)^*(\U^T\OO_T) = \OO_T^*\OO_T$.

Multiplying $\S_T^*\S$ out we obtain
\begin{equation*}
  \frac1M\Tr\left(\S_T^*\S_T\right) = \frac1M \sum_{s,f=1}^M |(\S_T)_{s,f}|^2.
\end{equation*}

We can expand the entries of $\S_T$ in terms of trajectories on the graph.
Using the definition of $\S_T$, we obtain
\begin{eqnarray*}
  (\S_T)_{sf} &=& \frac1T\sum_{t=0}^{T-1} \sum_{b_0,\ldots b_T}
  U_{b_0,b_1} \cdots U_{b_{T-t-1},b_{T-t}} O_{b_{T-t},b_{T-t}}
  \cdots U_{b_{T-1},b_T} \\
  &=&  \sum_{b_0,\ldots b_T} U_{b_0,b_1} \cdots U_{b_{T-1},b_T}
  \left(\frac1T\sum_{t=0}^{T-1} O_{b_{T-t},b_{T-t}} \right),
\end{eqnarray*}
where the inner sum in the first line is over all sequences of bonds
satisfying $b_0=s$ and $b_T=f$.  Such a sequence of bonds we will call a {\em
  trajectory}.  Only trajectories compatible with the graph's geometry (i.e.
those for which $U_{b_jb_{j+1}}\neq0$) contribute to $K(n,T)$.  A trajectory
$\tau = (b_0,\ldots,b_T)$ is said to have length $T$ and amplitude
\begin{equation*}
  A_\tau \mdef U_{b_0b_1} \cdots U_{b_{T-1}b_T}.
\end{equation*}
We will denote by $\Obs_{\tau}$ the average of the observable over the
trajectory $\tau$,
\begin{eqnarray*}
  \Obs_{\tau} \mdef \frac1T \left(O_{b_1b_1} + \ldots + O_{b_Tb_T}\right).
\end{eqnarray*}
To summarize, we have shown that
\begin{multline}
  \label{eq:trace_formula}
  V_n \leq K(n,T) = \frac1{M_n}\Tr\left(\S_T^*\S_T\right)
  = \frac1{M_n} \sum_{s,f=1}^{M_n} \left| \sum_{\tau:s\to f}
    \Obs_{\tau} A_\tau \right|^2 \\
  = \frac1{M_n} \sum_{s,f=1}^{M_n} \sum_{\tau_1, \tau_2:s\to f}
    \Obs_{\tau_1}^* \Obs_{\tau_2} A^*_{\tau_1} A_{\tau_2}.
\end{multline}
where the inner sum is over all possible trajectories of length $T$ starting
at $s$ and finishing at $f$.

%%%%%%%%%%%%%%%%%%%%%%%%%% New Section %%%%%%%%%%%%%%%%%%%%%%%%%%%
\section{Diagonal terms}

Equation (\ref{eq:trace_formula}) is reminiscent of a trace formula expansion
of the spectral form factor (i.e.~of the Fourier transform of the spectral
two-point correlation function)\footnote{To underline this similarity we have
  used in (\ref{eq:K_def}) the traditional notation for the form factor, $K$},
in particular of a graph, see {\em e.g.}~\cite{KS99}.  Such expansions are
notoriously difficult to analyze rigorously as both $T$ and the size of the
graph increase.  The starting point of any such analysis is the evaluation of
the contribution from the {\em diagonal terms\/}, obtained by restricting the
last sum in (\ref{eq:trace_formula}) to identical trajectories,
$\tau_1=\tau_2$.  It is usually assumed that the off-diagonal terms sum up to
a subdominant contribution, when $T$ and the size of the graph scale
appropriately.  This idea, called the {\em diagonal approximation\/} was first
introduced for a general class of systems in \cite{Ber85}.  On graphs it was
explored, in particular, in \cite{KS99,Tan01}.  It is difficult,
however, to give an a priori estimate on the size of the off-diagonal
contributions and the analysis is usually restricted to evaluating the
contributions coming from specific classes of interacting trajectories
\cite{BSW02,BSW03,B04,B06}.

Our strategy now is to calculate the contribution from the diagonal terms in
(\ref{eq:trace_formula}).  Then we will show that, in the
case of graphs constructed from 1d maps, we can actually estimate the
off-diagonal terms by virtue of being able to choose an appropriate $T=T(n)$.

To evaluate the diagonal contribution
\begin{equation*}
  \Kdiag \mdef 
  \frac1{M_n} \sum_{\tau} \left|\Obs_\tau\right|^2 \left|A_\tau\right|^2,
\end{equation*}
we make two observations.  First, by the definition of the amplitude $A_\tau$
and the defining property of the matrix $\U$, equation (\ref{eq:U_def}), we
obtain
\begin{equation*}
  \left|A_\tau\right|^2 = |U_{b_0,b_1}|^2\cdots |U_{b_{T-1},b_T}|^2
  = B_{b_0,b_1} \cdots B_{b_{T-1},b_T}.
\end{equation*}
Now we recall Lemma~\ref{lem:B_properties},
part~\ref{item:prod_B_measure}, and conclude that
\begin{equation*}
  \left|A_\tau\right|^2
  = \frac{\mu\left( \bigcap_{t=0}^T S^{-t}(E_{b_t}) \right)}
  {\mu\left(E_{b_0}\right)}.
\end{equation*}
On the other hand, by definition of $\Obs_\tau$,
\begin{equation*}
  \Obs_\tau = \frac1T \sum_{t=1}^T O_{b_t,b_t}
  = \frac1T \sum_{t=1}^T \wh{\obs}(E_{b_t})
  = \frac1T \sum_{t=1}^T \wh{\obs}\circ S^t\left(S^{-t}(E_{b_t})\right),
\end{equation*}
where $\wh{\obs}(E_b)$ denotes the (constant) value of the function
$\wh{\obs}$ on the atom $E_b$.  In fact, it is easy to see that $\Obs_\tau$
coincides with the value of the function
\begin{equation*}
  \wh{\obs}_T \mdef \frac1T\sum_{t=1}^T\wh{\obs}\circ S^t
\end{equation*}
on the set $\bigcap_{t=0}^T S^{-t}(E_{b_t}) \mdef E_{b_0,\ldots,b_T}$, if
this set is non-empty.  If it is empty, the value of $\Obs_\tau$ is of no
consequence since the trajectory $\tau$ is then incompatible with the graph's
geometry and $A_\tau=0$.

The measure of all atoms $E_b$ is assumed to be equal.  More precisely, it is
equal to $1/M_n$, since $M_n$ is the total number of the atoms.  Collecting our
observations together, we can express the diagonal term as
\begin{eqnarray*}
  \Kdiag
  &=& \frac1{M_n} \sum_{\tau} 
  \left(\wh{\obs}_T\left(E_{b_0,\ldots,b_T}\right)\right)^2 
  \frac{\mu(E_{b_0,\ldots,b_T})}{M_n^{-1}}\\
  &=& \int_0^1 \left(\wh{\obs}_T(x)\right)^2 dx
  = \E\left(\frac1T\sum_{t=1}^T\wh{\obs}\circ S^t\right)^2 
  = \cV_T(\wh{\obs}).
\end{eqnarray*}
Thus, by the $L^2$ ergodic theorem (Theorem~\ref{thm:L2ergodic}), $\Kdiag$
goes to zero as $T\to\infty$.  On the other hand, $K(n,T)$ is bounded below by
a non-negative $V_n$ which is, generically, non-zero for a fixed $n$.  This
shows that the diagonal term is a poor approximation to $K(n,T)$ in the limit
$T\to\infty$.  Luckily, this is not the limit we have to take.

%%%%%%%%%%%%%%%%%%%%%%%%%%%%%% New Section %%%%%%%%%%%%%%%%%%%%%%%%%%%%%
\section{Completion of the proof of quantum ergodicity}

Lemma~\ref{lem:traj_spec} has a very important consequence for the inner sum
in (\ref{eq:trace_formula}).
\begin{lemma}
  \label{lem:diag_exact}
  The diagonal term $\Kdiag$ gives the exact value of
  $K(n,T)$ up to time $T=n$, {\em i.e.}
  \begin{equation*}
    K(n,T) = \Kdiag = \cV_T(\wh{\obs})
    \qquad \mbox{if }T\leq n,
  \end{equation*}
  where $\wh{\obs} = \condex{\obs}{\cM_n}$.
\end{lemma}

\begin{proof}
  By Lemma~\ref{lem:traj_spec}, for every pair of bonds $s$ and $f$, there is
  at most one trajectory going from $s$ to $f$ in $T\leq n$ steps.
  Thus, for $T\leq n$,
  \begin{multline*}
    K(n,T) = \frac1{M_n} \sum_{s,f=1}^{M_n} \left| \sum_{\tau:s\to f}
      \Obs_{\tau} A_\tau \right|^2 \\
    = \frac1{M_n} \sum_{s,f=1}^{M_n}
    \left|\Obs_{\tau(s\to f)} A_{\tau(s\to f)}\right|^2
    = \frac1{M_n} \sum_\tau \left|\Obs_\tau A_\tau\right|^2 = \Kdiag.
  \end{multline*}
\end{proof}

As a consequence, we have the following result.

\begin{theorem} {\bf(Quantum Ergodicity)}
  \label{thm:variance_0}
  Let the map $S$ and the sequence of partitions $\{\cM_n\}$ satisfy
  Conditions \ref{assum:map}, \ref{assum:partition} and \ref{assum:unistoch};
  let $\{\U_n\}$ be the corresponding sequence of unitary matrices with
  eigenvectors $\evU^{(n)}_j$; and let $\{\OO_n\}$ be a sequence of diagonal
  matrices corresponding to an observable $\obs\in L^2[0,1]$ with $\E\obs=0$.
  If $S$ is ergodic, then
  \begin{equation*}
    V_n = \frac1{M_n} \sum_{j=1}^{M_n}
    \left| \left(\evU^{(n)}_j, \OO_n \evU^{(n)}_j\right) \right|^2 \to 0
    \qquad \mbox{as } n\to\infty.
  \end{equation*}
\end{theorem}

\begin{proof}
  The variance $V_n$ is majorized by $K(n,T)$ for any $T$.  We combine
  Lemma~\ref{lem:diag_exact} with equation (\ref{eq:local_av_conv}) we
  conclude that, for a fixed $T$,
  \begin{equation*}
    K(n,T) \to \ergvar_T(\obs)\qquad \mbox{as }n\to\infty.
  \end{equation*}
  Now we use the standard $\eps/2$ argument: for any $\eps>0$, by
  Theorem~\ref{thm:L2ergodic} we can find $T$ such that
  $\ergvar_T(\obs)<\eps/2$.  Having fixed this $T$, we find $n(\eps,T)$ such
  that $|K(n,T) - \ergvar_T(\obs)|<\eps/2$ for all $n\geq n(\eps,T)$.
  Combining the above,
  \begin{equation*}
    V_n \leq K(n,T) < \eps/2+\eps/2
  \end{equation*}
  as long as $n\geq n(\eps,T)$.  Since $\eps$ was arbitrary, we conclude that
  $V_n\to0$.
\end{proof}

\begin{remark}
  One can avoid the $\eps/2$ argument in the following way.  Taking the limit
  $n\to\infty$ of the inequality $V_n \leq K(n,T)$ produces
  \begin{equation*}
    0 \leq \limsup_{n\to\infty} V_n \leq \limsup_{n\to\infty} K(n,T) 
    = \ergvar_T(\obs).
  \end{equation*}
  Now taking the $T\to\infty$ limit, we obtain
  \begin{equation*}
    0 \leq \limsup_{n\to\infty} V_n 
    = \limsup_{T\to\infty} \limsup_{n\to\infty} V_n 
    \leq \limsup_{T\to\infty} \ergvar_T(\obs) = 0.
  \end{equation*}
\end{remark}

%%%%%%%%%%%%%%%%%%%%%%%%%%%%%% New Section %%%%%%%%%%%%%%%%%%%%%%%%%%%%%
\section{Egorov property}

In Section~\ref{sec:quant_observable} we defined a procedure to obtain a
piecewise constant function $\wh{\obs}$ given a function $\obs\in L^2[0,1]$.
It is enlightening to see how $\wh{\obs\circ S}$ is related to $\wh{\obs}$.

By definition,
\begin{equation*}
  \wh{\obs\circ S}\Big|_{E_j} = \frac1{\mu(E_j)} \int_{E_j} \obs(S(y)) d\mu(y)
  = \frac1{\mu(E_j)} \int_{S(E_j)} \obs(z) \frac{d\mu(z)}{|S'(y)|}.
\end{equation*}
Since $S$ is linear on $E_j$, its derivative is constant.  In
fact, it is easy to see that $1/|S'(y)| = B_{jk}$, where $y\in
E_j$ and $S(y)\in E_k$.  Thus we have
\begin{equation*}
  \wh{\obs\circ S}\Big|_{E_j} = \sum_{k: E_k \cap S(E_j) \neq \emptyset}
  B_{jk} \frac1{\mu(E_j)} \int_{E_k} \obs(z) d\mu(z),
\end{equation*}
where the sum is over the decomposition of the set $S(E_j)$ into atoms $E_k$.
Since $B_{jk}=0$ whenever $E_k \cap S(E_j)$ is empty and since $\mu(E_j)$ is
independent of $j$, we arrive to the following conclusion
\begin{lemma}
  \label{lem:discr_f_of_S}
  If, for a given partition $\cM$, the matrices $\B$, $\OO(\obs)$ and
  $\OO(\obs\circ S)$ are defined according to (\ref{eq:markov_def}) and
  (\ref{eq:O_def}) then
  \begin{equation*}
    O_{jj}(\obs\circ S) = \sum_{k=1}^{M} B_{jk} O_{kk}(\obs),
  \end{equation*}
  where $M$ is the number of atoms in the partition $\cM$.
\end{lemma}

Lemma~\ref{lem:discr_f_of_S} is a rather beautiful manifestation of the
inter-consistency between the discretization procedures for maps $S$ and
observables $\phi\in L^2$.  Namely, the discretization commutes with the
action of $S$ on $L^2$.  In this, Lemma~\ref{lem:discr_f_of_S} is a classical
analogue of the Egorov property, a result which shows that the unitary
matrices $\U_n$ faithfully represent the action of the classical map $S$.

\begin{theorem}{\bf (Egorov property)}
  \label{thm:egorov}
  Let the map $S$ and the sequence of partitions $\{\cM_n\}$ satisfy
  Conditions \ref{assum:map}, \ref{assum:partition} and \ref{assum:unistoch};
  let $\{\U_n\}$ be the corresponding sequence of unitary matrices with
  eigenvectors $\evU^{(n)}_j$; and let $\{\OO_n\}$ be a sequence of diagonal
  matrices corresponding to an observable $\obs$.  If $\obs$ is Lipschitz
  continuous then
  \begin{equation*}
    \| \U_n\OO_n(\obs)\U_n^{-1} - \OO_n(\obs\circ S) \| = O({M_n}^{-1}),
  \end{equation*}
  where the norm is the operator norm on the Euclidean space $\Reals^{M_n}$.
\end{theorem}

\begin{proof}
  We fix the partition $\cM_n$, denote the corresponding $\U\OO\U^{-1}$ by
  $\Q$ and observe that, while $\OO(\obs\circ S)$ is a diagonal matrix, $\Q$
  is not necessarily so.

  First we treat the diagonal elements of $\Q$.  Writing them out explicitly
  we get
  \begin{equation*}
    Q_{jj} = \sum_{r=1}^{M_n} U_{jr} O_{rr} U^{-1}_{rj}
    = \sum_{r=1}^{M_n} U_{jr} O_{rr} \wb{U_{jr}}
    = \sum_{r=1}^{M_n} |U_{jr}|^2 O_{rr}
    = O_{jj}(\obs\circ S),
  \end{equation*}
  where we used the unitarity of $\U$ and its defining property, $|U_{jr}|^2 =
  B_{jr}$ and Lemma~\ref{lem:discr_f_of_S}.

  For the off-diagonal elements of $\Q$ we have
  \begin{eqnarray*}
    Q_{jk} = \sum_{r=0}^{M_n} U_{jr} O_{rr} U^{-1}_{rk}
    &=& \sum_{r=0}^{M_n} U_{jr} (O_{rr}-C) \wb{U_{kr}}
    + C \sum_{r=0}^{M_n} U_{jr} \wb{U_{kr}} \\
    &=& \sum_{r=0}^{M_n} U_{jr} (O_{rr}-C) \wb{U_{kr}},
  \end{eqnarray*}
  where $C$ is any constant and we have used the unitarity of $\U$ to conclude
  that the second sum is zero.  We estimate, using Cauchy-Schwarz,
  \begin{eqnarray*}
    |Q_{jk}| &\leq& \max_r |O_{rr}-C| \sum_{r=0}^{M_n} |U_{jr} \wb{U_{kr}}|\\
    &\leq& \max_r |O_{rr}-C|
    = \max_{x\in S(E_j)\cap S(E_k)}  |\wh{\obs}(x)-C|.
  \end{eqnarray*}
  If $\obs$ is Lipschitz continuous with
  \begin{equation*}
    \| \obs \|_{\mathrm{Lip}}
    \mdef \sup_{x\neq y}\frac{|\obs(x)-\obs(y)|}{\mu(x,y)} < \infty,
  \end{equation*}
  we can estimate further, by choosing appropriate $C$,
  \begin{equation*}
    |Q_{jk}| \leq \frac12 \big\| \wh{\obs} \big\|_{\mathrm{Lip}}
    \mu\big(S(E_j)\cap S(E_k)\big)
    \leq \frac12 \| \obs \|_{\mathrm{Lip}} \max|S'(x)| \mu(E_j)
    \propto \frac{\| \obs \|_{\mathrm{Lip}}}{M_n}.
  \end{equation*}
  
  Since $U_{jr}$ is non-zero only if $S(E_j) \cap E_r \neq \emptyset$ and
  $U_{kr}$ is non-zero only if $S(E_k) \cap E_r \neq \emptyset$, we conclude
  that $Q_{jk}=0$ if $S(E_j)$ and $S(E_k)$ are disjoint.  Thus the matrix $\Q$
  is of block-diagonal structure, each block corresponding to an equivalence
  class as defined by Lemma~\ref{lem:intersect_classes}.  The norm of $\Q
  - \OO(\obs\circ S)$ is equal to the maximum of the norms of the blocks.  A
  norm of a block, in turn, is bounded by its dimension times the maximum
  absolute value of the element of the block.  The dimension of a block is
  uniformly bounded by Lemma~\ref{lem:intersect_classes}.  Thus we get
  \begin{equation*}
    \| \Q - \OO(\obs\circ S) \|
    \leq D(S) \frac{\| \obs \|_{\mathrm{Lip}}}{M_n}
  \end{equation*}
  for some constant $D(S)$ which is independent of $\obs$ and $n$.
\end{proof}

\begin{remark}
  If the function $\obs$ is only assumed to be continuous on $[0,1]$,
  one can prove a weaker property:
  \begin{equation*}
    \| \U_n\OO_n(\obs)\U_n^{-1} - \OO_n(\obs\circ S) \| \to 0
    \qquad \mbox{as } M_n\to\infty.
  \end{equation*}
\end{remark}

%%%%%%%%%%%%%%%%%%%%%%%%%%%%%% New Section %%%%%%%%%%%%%%%%%%%%%%%%%%%%%
\section{Discussion}

We have succeeded in proving quantum ergodicity (QE) for a special class of
sequences of quantum graphs.  However, we would like to mention that the
result is expected to hold for much broader class of graphs.

It is true that, given a finite quantum graph $G$, one can associate a 1d map
to it by reversing the process described in the paper.  Thereafter, it is
possible to produce a sequence of graphs, one of which will coincide with the
original graph $G$, and answer the question of QE for this sequence.  In this
sense, each graph corresponds to a 1d map.  However, this is not true for
every {\em sequence\/} of graphs.  In fact, it is not true for most sequences.
Examples of such sequences include star graphs with Kirchhoff conditions at
the central vertex (for which the question of QE has been answered
negatively), the complete (Kirchhoff) graphs, and the star graphs with Fourier
central vertex \cite{Tan01}, for both of which the QE is expected (but is not
known) to hold in some form.

It is reassuring that the proof of QE in the present article suggests a
direction for possible generalizations: study the diagonal terms and then find
an estimate for the off-diagonal ones.  However, for the sequences of graphs
described above the diagonal approximation ceases to be exact for $T>1$ (cf.
Lemma~\ref{lem:diag_exact}).  This makes estimation of the off-diagonal terms
a much more difficult task.

Another interesting question to consider is whether quantum {\em unique\/}
ergodicity (when the convergence in (\ref{eq:conv_to_mean}) happens along {\em
  all\/} sequences of eigenvectors) is true for any quantum graphs.  This has
been answered in the negative \cite{SK03} for graphs with Kirchhoff vertices
but is unclear for other types if graphs.

%%%%%%%%%%%%%%%%%%%%%%%%%%%%%% Acknowledgement %%%%%%%%%%%%%%%%%%%%%%%%%%%
\section*{Acknowledgement}

We would like to thank Zeev Rudnick for his suggestion to consider proving
Egorov property, which we followed with success.  We are also grateful to
Alexander G.\ Kachurovskii for enlightening discussions on the speed of
convergence in ergodic theorems.

One of the authors (GB) wishes to thank the University of Bristol and the
Weizmann Institute of Science for the hospitality extended to him.

This collaboration was supported by EPSRC Grant GR/T06872/01.  JPK is
supported by an EPSRC Senior Research Fellowship.

%%%%%%%%%%%%%%%%%%%%%%%%%%%%%% Appendix %%%%%%%%%%%%%%%%%%%%%%%%%%%%%%%%%%
\appendix
\section{Connection between variances $V^S$ and $V^U$}
\label{sec:var_rel}

To demonstrate relation (\ref{eq:var_rel}) we start with summarizing the
notation introduced in Section~\ref{sec:q_graphs}.  Let the unitary $B\times
B$ matrix $\S$ be defined by $\S = \S(\spec) = e^{i\spec\L}\S(0)$, where $\L$
is the diagonal matrix of the bond lengths of the graph and $\S(0)$ is some
fixed unitary matrix.  Let $\{\spec_n\}$ be the (real) solutions of the
equation $\det (\I - \S(\spec) ) = 0$.  We assume that the spectrum
$\{\spec_n\}$ is non-degenerate, which is a generic situation
\cite{Fri05}.

Denote by $\evS_n$ the normalized eigenvector of $\S(\spec_n)$ corresponding
to the eigenvalue 1.  By $\evU_k(\spec)$ we denote the $k$-th normalized
eigenvector of $\S(\spec)$.  We further denote by $e^{i\theta_k(\spec)}$ the
eigenvalues of $\S(\spec)$, with $\theta_k$ chosen to be continuous (indeed
smooth) functions of $\spec$.

When $\spec=\spec_n$ there is an index $k$ for which $\theta_k(\spec) = 0 \mod
2\pi$.  For this index $k$ we also have $\evS_n = \evU_k(\spec)$.

Let $\A$ be a self-adjoint matrix (a generalization of the observable $\OO$)
with trace 0 (without loss of generality).  We are interested in the
relationship between two variances,
\begin{equation*}
  V^S(\Spec,B) = \frac1{\overline{N}(\Spec)} \sum_{\spec_n \leq \Spec}
  \langle \evS_n|\A|\evS_n\rangle^2,
\end{equation*}
and
\begin{equation*}
  V^U(\S(\spec),B) = 
  \frac1{B} \sum_{k=1}^B \langle \evU_k(\spec)|\A|\evU_k(\spec)\rangle^2,
\end{equation*}
where $\overline{N}(\Spec) = \Spec \Tr\L / 2\pi$ is the mean number of the
eigenvalues $\spec_n$ in the interval $[0,\Spec]$.

Introducing the notation
\begin{equation*}
  A_k = A_k(\spec) = \left(\evU_k(\spec), \A\evU_k(\spec)\right)
\end{equation*}
we observe that
\begin{equation}
  \label{eq:trace_expansion}
  \Tr \A\S^m(\spec) = \sum_{k=1}^B A_k e^{im\theta_k(\spec)}.
\end{equation}
In particular, $\Tr \A = \sum_{k=1}^B A_k = 0$.
From the theory of distributions we know that
\begin{equation*}
  \Delta_\eps(\theta) \mdef \frac1{2\pi} \left(1+
    \sum_{m=1}^\infty e^{-m\eps}\left(e^{im\theta}+e^{-im\theta}\right)
  \right)
\end{equation*}
converges, in the limit $\eps\to0$, to
\begin{equation*}
  \Delta(\theta) \mdef \sum_{r=-\infty}^\infty \delta(\theta - 2\pi r) ,
\end{equation*}
where $\delta$ is the Dirac delta function.  Substituting in the above
identity $\theta=\theta_k$, multiplying by $A_k$ and performing the summation
over $k$ yields
\begin{equation}
  \label{eq:trace_form_A}
  \sum_{k=1}^B \Delta_\eps(\theta_k) A_k = \frac1{2\pi} \left(
    \sum_{m=1}^\infty e^{-m\eps}\left(\Tr\A\S^m + \Tr\A\S^{-m}\right)
  \right).
\end{equation}
As $\eps\to0$ this converges to 
\begin{equation*}
  \sum_{k=1}^B \Delta(\theta_k) A_k 
  = \sum_{n=1}^\infty \frac{A_k(\spec_n)}{|\theta_k'(\spec_n)|} 
  \delta(\spec-\spec_n),
\end{equation*}
where, given $\spec_n$, $k$ is chosen to satisfy $\theta_k=0$.  It is shown in
\cite{KS99} that
\begin{equation*}
  \theta_k'(\spec_n) = \left(\evS_n, \L\evS_n\right)
  = \big(\evU_k(\spec_n), \L\evU_k(\spec_n)\big) \mdef L_k(\spec_n).
\end{equation*}
Clearly, $\theta_k'(\spec_n)>0$ and so we can drop the modulus around
$\theta_k'(\spec_n)$ in the previous equation.  Now we need the following
properties of the approximants to the Dirac delta function,
\begin{equation*}
  2\pi \lim_{\eps\to0} \eps \Delta_\eps^2(x) = \Delta(x)
\end{equation*}
and, if $x_1\neq x_2$, 
\begin{equation*}
  2\pi \lim_{\eps\to0} \eps 
  \left(\Delta_\eps(x-x_1) + \Delta_\eps(x-x_2)\right)^2 
  = \Delta(x-x_1) + \Delta(x-x_2).
\end{equation*}
Applying these identities gives
\begin{equation*}
  2\pi \lim_{\eps\to0} \eps 
  \left(\sum_{k=1}^B \Delta_\eps(\theta_k) A_k \right)^2
  = \sum_{k=1}^B \Delta(\theta_k) A_k^2 
  = \sum_{n=1}^\infty \frac{A_k^2(\spec_n)}{L_k(\spec_n)} 
  \delta(\spec-\spec_n).
\end{equation*}
Integrating the right-hand side with respect to $\spec$ we get
\begin{equation*}
  \frac1{\overline{N}(\Spec)} 
  \int_{0}^\Spec \left(
    \sum_{n=1}^\infty \frac{A_k^2(\spec_n)}{L_k(\spec_n)} 
    \delta(\spec-\spec_n) \right) d\spec
  = \frac1{\overline{N}(\Spec)} \sum_{\spec_n<\Spec} 
  \frac{A_k^2(\spec_n)}{L_k(\spec_n)} \mdef \widehat{V}^S(\Spec,B)
\end{equation*}
We will use this quantity to approximate $V^S(\Spec,B)$.  It is a good
approximation if the bond lengths of the graph are approximately 1 (i.e. the
matrix $\L$ is approximately unity):
\begin{equation*}
  L_{\min} \widehat{V}^S(\Spec,B) 
  \leq V^S(\Spec,B) \leq L_{\max} \widehat{V}^S(\Spec,B),
\end{equation*}
where $L_{\max}$ and $L_{\min}$ are the maximal and minimal bond lengths.

Using (\ref{eq:trace_form_A}) and expanding the square we obtain
\begin{eqnarray*}
  \widehat{V}^S(\Spec,B) &=& \frac1{\overline{N}(\Spec)}
  \int_{0}^\Spec 2\pi \lim_{\eps\to0} \eps 
  \left(\sum_{k=1}^B \Delta_\eps(\theta_k) A_k \right)^2 d\spec\\
  &=& \frac1{2\pi \overline{N}(\Spec)}
   \int_{0}^\Spec \lim_{\eps\to0} \eps
  \!\!\!\sum_{m_1, m_2=-\infty}^\infty \!\!
  e^{-\eps(|m_1|+|m_2|)} \Tr\A\S^{m_1} \Tr\A\S^{m_2} d\spec.
\end{eqnarray*}
We now take the $\Spec\to\infty$ limit of the above expression and interchange
it with the $\eps$-limit.  Due to the rational independence of the bond
lengths, the limit
\begin{equation*}
  \lim_{\Spec\to\infty} \frac{1}{\overline{N}(\Spec)} 
  \int_0^\Spec \Tr\A\S^{m_1} \Tr\A\S^{m_2} d\spec
\end{equation*}
is zero whenever $m_1 \neq -m_2$.  Indeed, we recall that $\S = \S(\spec) =
e^{i\spec\L}\S(0)$ and expand the trace
\begin{equation*}
  \Tr\A\S^{m} = \sum_{b_0,\ldots,b_m} 
  A_{b_0,b_1} S(0)_{b_1,b_2} \cdots S(0)_{b_m,b_0} \exp(i\spec L_p),
\end{equation*}
where $L_p = L_{b_1}+\cdots+L_{b_m}$.  If $L_p$ is a sum of $m_1$ terms and
$L_q$ is a sum of $m_2\neq m_1$ terms, they cannot be equal.  Thus, the only
case when the phase factors $\exp(i\spec L_p)$ in a product of two traces
can cancel each other is when $m_1=-m_2$.

We arrive at
\begin{eqnarray*}
  \lim_{\Spec\to\infty} \widehat{V}^S(\Spec,B) 
  &=& \lim_{\Spec\to\infty} \frac1{\pi \overline{N}(\Spec)} 
  \int_{0}^\Spec  \lim _{\eps \rightarrow 0} \eps
  \sum_{m=1}^{\infty}  e^{-2m\eps} \Tr\A\S^m \Tr\A\S^{-m} \,d\spec \\
  &=&  \lim_{\Spec\to\infty} \frac1{2\pi\overline{N}(\Spec)} 
  \int_{0}^\Spec \sum_{k=1}^B A_k^2(\spec) \,d\spec \\
  &=& \lim_{\Spec\to\infty} \frac1{2\pi\overline{N}(\Spec)} 
  \int_{0}^\Spec V^U(\S(\spec),B) \,d\spec.
\end{eqnarray*}
Here we used expansion (\ref{eq:trace_expansion}) and the fact that
\begin{equation*}
  \lim _{\eps \rightarrow 0} \eps
  \sum_{m=1}^{\infty} e^{-2m\eps} e^{im(\theta_{k_1}-\theta_{k_2})}
  = \lim _{\eps \rightarrow 0}
  \frac{\eps}{e^{2\eps-i(\theta_{k_1}-\theta_{k_2})} - 1}
\end{equation*}
is $1/2$ when $\theta_{k_1}=\theta_{k_2}$ and 0 otherwise.

%\bibliographystyle{ieeetr}
%\bibliography{all,berkolaiko}

\def\cprime{$'$}

\end{document}